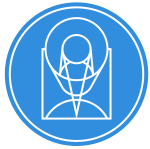

# JWST TECHNICAL REPORT

| Title: Global sky background images for JWST/NIRISS Wide-Field Slitless Spectroscopy | Doc #: JWST-STScI-009057, SM-12<br>Date: 16 June 2025<br>Rev: |
|---|---|
| Authors: Gaël Noirot and the NIRISS WFSS team    Phone: (410) 338-4763 | Release Date: 10 July 2025 |

## 1. Abstract

We present updated empirical background images for JWST/NIRISS Wide-Field Slitless Spectroscopy (WFSS) for both orthogonal grisms (GR150C, GR150R) crossed with all NIRISS WFSS wide-band filters (F090W, F115W, F150W, F200W). The background images are created using carefully vetted science and calibration exposures and improve the quality of previous background reference files. We present our methodology to create the background images and assess their quality using background subtracted science data. Overall, background residuals reach below 1% of the sky brightness for all filter and grism combinations, corresponding to a decrease in the background RMS of a factor of up to 7 from previous background reference files. The new background images are available on the Calibration Reference Data System (CRDS[1]) as of context 1365 and we recommend using them for improved quality of NIRISS WFSS data.

## 2. Introduction

The Wide-Field Slitless Spectroscopy (WFSS, Willott et al. 2022) mode of observation of the Near-InfraRed Imager and Slitless Spectrograph (NIRISS, Doyon et al. 2023) onboard JWST (Gardner et al. 2023) enables the acquisition of up to thousands of spectra per exposure without prior target preselection. In WFSS mode, JWST/NIRISS offers a 2.2 x 2.2 arcmin$^2$ field of view, with two, low-resolution (R~150), orthogonal grisms (GR150C and GR150R) that can be crossed with four wide-band (F090W, F115W, F150W, F200W) and two medium-band (F140M, F158M) near-IR filters covering the 8000-22000 Å wavelength range.

Prior to JWST, this observing mode has proven highly successful with the ACS G800L, WFC3/UVIS G280, WFC3/IR G102 and G141 grisms onboard the Hubble Space Telescope (HST), enabling transformative science from the Milky-Way to the highest redshifts (e.g., Pirzkal et al. 2004, Atek et al. 2010, Brammer et al. 2012, Newman et al. 2014, Buenzli et al. 2015, Momcheva et al. 2016, Nelson et al. 2016, Oesch et al. 2016,

---

[1] https://jwst-crds.stsci.edu

Noirot et al. 2016, 2018, 2022, Lotz et al. 2017, Estrada-Carpenter et al. 2019, 2023, Matharu et al. 2019, 2021, 2022, Wang et al. 2020, D'Eugenio et al. 2021, Noirot & Sawicki 2022, Seymour et al. 2022, 2024). Thanks to JWST's exceptional sensitivity and spatial resolution at near-IR wavelengths, JWST/NIRISS WFSS has already demonstrated its capabilities with science highlights from the nearby Universe to the epoch of reionization (e.g., Boyett et al. 2022, Roberts-Borsani et al. 2022, Noirot et al. 2023, Bradač et al. 2024, Rihtaršič et al. 2025, Shen et al. 2025), and is poised to further our understanding of galaxy formation and evolution, notably into the regime of spatially resolved studies of galaxy populations from late to early cosmic epochs (e.g., Mowla et al. 2022, Wang et al. 2022, Matharu et al. 2023, Estrada-Carpenter et al. 2024, Shen et al. 2024). The next generation of wide-field space-based telescopes such as Euclid (Euclid Collaboration et al. 2024), the Nancy Grace Roman Space Telescope (Roman; Akeson et al. 2019), and the Cosmological Advanced Survey Telescope for Optical and ultraviolet Research (CASTOR; Côté et al. 2025), which are all equipped with WFSS capabilities, further demonstrate the versatility and power of this observing mode, over field of views 100x greater than that of JWST.

By design, JWST/NIRISS WFSS disperses all the light entering the pupil and filter wheels, including that of the sky background which is dispersed over the entire detector. Because of the finite size of the entrance pupil, the different spectral orders of the sky background are dispersed over finite lengths in the dispersion direction and, depending on orders, do not contribute at certain positions on the detector (Kümmel et al. 2011). This imprints specific profiles of varying intensity along the dispersion direction in WFSS images, whose overall shapes are different depending on filter and grism combinations. Accurate modelling and removal of the background light is therefore critical for the proper extraction and analysis of WFSS data, especially that of fainter spectra. Here, we present updated empirical background images for NIRISS WFSS created using publicly available data taken between 2022 May 05 and 2024 June 11, which improve the quality of background subtracted science images compared to previous WFSS background reference files.

The structure is as follows. Section 3 describes our methodology to create the background images for all NIRISS wide-band filter and grism combinations. Section 4 assesses the quality of the images, and we summarize our results in Section 5.

## 3. Methodology

### 3.1. Initial WFSS backgrounds

The left panel of Figure 1 shows an example JWST/NIRISS WFSS science image taken with the GR150C grism crossed with the F200W filter. The varying intensity of the background profile is clearly visible as slightly-tilted vertical bands over the entire detector. The top panel shows the median background profile along the dispersion direction, derived from regions free of spectral traces. The sharpness of the transitions seen in the median profile is affected by the slight tilt of the vertical bands. This implies a small tilt between the pick-off mirror edges as projected on the pupil wheel plane and the detector orientation within the instrument. The right panel of the Figure shows the median background profiles along the dispersion direction for example long-exposure pointings in all wide-band filters (F090W, F115W, F150W, F200W) crossed with the GR150C (black solid lines) and GR150R (grey solid lines) grisms. The different shapes and varying intensities of the background profile in the various filter and grism combinations are clearly visible. While the background profiles are primarily due to the overlap of the different spectral orders of the sky background, additional features include a horizontal band of scattered light called the "lightsaber" (e.g., Rigby et al. 2023a), and the effect of the pick-off mirror coronographic spots (e.g., Willott et al. 2022), which may be seen as additional bumps and troughs at specific pixel positions in the 2D images or 1D profiles, respectively. The methodology we adopt to create 2D background models for each filter and grism combination is as follows:

- First, we query the Mikulski Archive for Space Telescopes (MAST[2]) and retrieve all public, calibrated NIRISS WFSS rate files available at the time of retrieval, for all wide-band filter and grism combinations. We then reject images taken in subarray mode, and exclude exposures with integrations shorter than 300 seconds to mitigate noisy sky backgrounds.

- We then perform a visual inspection of all WFSS images and remove exposures markedly affected by strong lightsabers, bright stars and diffraction spikes, and extended bright sources. We also reject extremely crowded star fields and flag crowded extragalactic fields. The latter are predominantly intermediate-redshift galaxy clusters comprised of numerous bright ellipticals and extended light which require a more careful treatment to mask their signal in the dispersed images compared to sparse fields devoid of such objects. Table 1 summarizes the number, date, and program IDs of the vetted exposures for each filter and grism combination. Figure 2 shows the median background level as a function of ecliptic latitude of our vetted and rejected exposures, highlighting the good sky coverage of our exposures, except for the F090W filter limited to a handful of pointings.

- Following this visual vetting, we mask bad pixels and reference pixels from the WFSS and corresponding flat-field images, and normalize the WFSS exposures with sigma-clipped medians. Similarly to the creation of HST WFC3/IR grism backgrounds (Kümmel et al. 2011, Brammer et al. 2012) or other NIRISS WFSS backgrounds from the literature (Hviding et al. 2024), we also apply the flat-field to remove flat-field

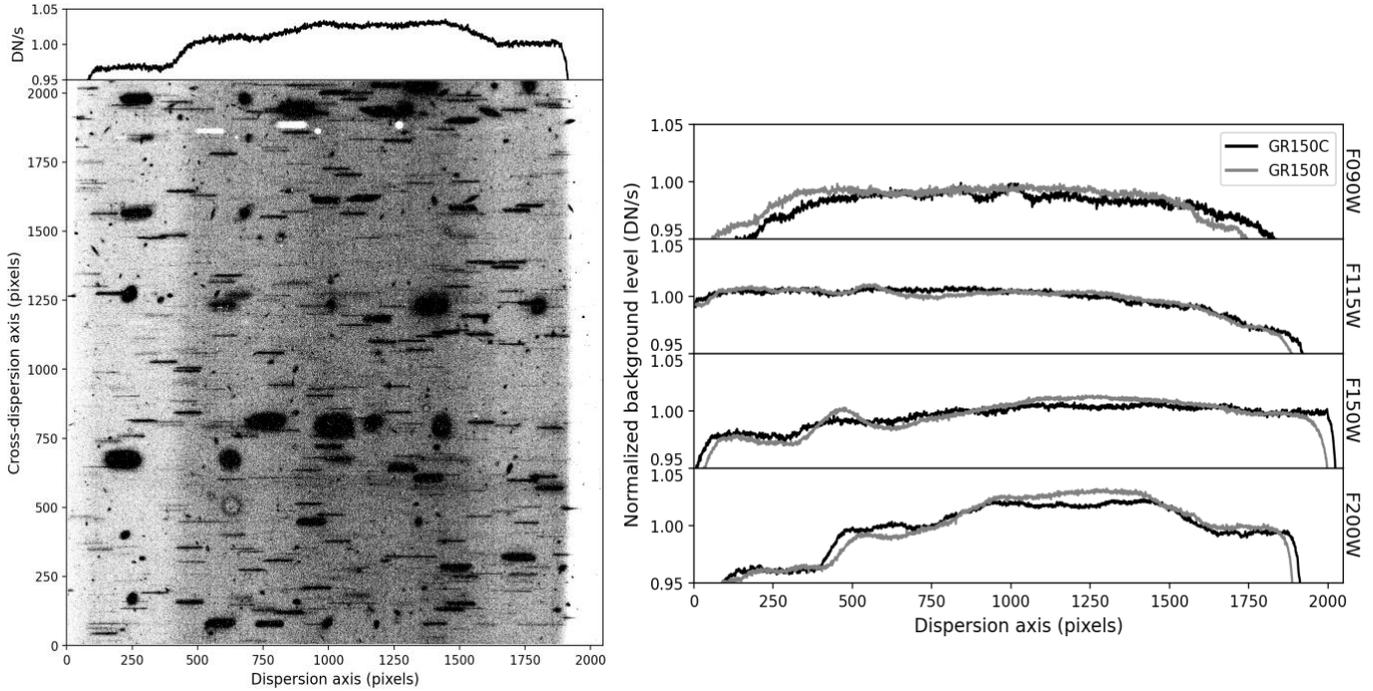

Figure 1. *Left*: Example F200W/GR150C science exposure (normalized). The top panel shows the normalized median 1D profile along the dispersion direction, after masking out the spectral traces. The varying background intensity along the dispersion axis is clearly seen in both panels. *Right*: The normalized median 1D profile of example science exposures for all wide-band filter and grism combinations, after trace masking. The GR150C (GR150R) profiles are shown in black (grey). These showcase the different shapes of the background across the different filter and grism combinations.

Table 1. Number, UT date of the last exposure taken, and program IDs of the vetted exposures for each filter and grism combinations. For each filter and grism pair, the UT date of the first exposure taken is 2022 May 05. For each filter, the program IDs are the same for both orthogonal grisms, except for program ID 4681 which only consists of GR150C exposures in our samples (denoted with a superscript C).

| Filter | Grism | # of exposures (vetted/inspected) | UT Date Last | Program IDs |
|--------|-------|-----------------------------------|--------------|-------------|
| F090W | GR150C | 36/67 (54%) | 2024/05/09 | 1089, 3362 |
|       | GR150R | 35/67 (52%) | 2024/05/08 | |
| F115W | GR150C | 156/258 (60%) | 2024/06/03 | 1089, 1208, 1283, 1324, 1571, |
|       | GR150R | 258/408 (63%) | 2024/06/01 | 2079, 3383, 4681[C] |
| F150W | GR150C | 201/333 (60%) | 2024/06/03 | 1089, 1208, 1283, 1324, 1571, |
|       | GR150R | 245/336 (73%) | 2024/06/01 | 2079, 3383, 4681[C] |
| F200W | GR150C | 276/448 (62%) | 2024/06/11 | 1089, 1208, 1283, 1324, 1571, |
|       | GR150R | 199/303 (66%) | 2024/05/16 | 2079, 2736, 2738, 3383, 4681[C] |

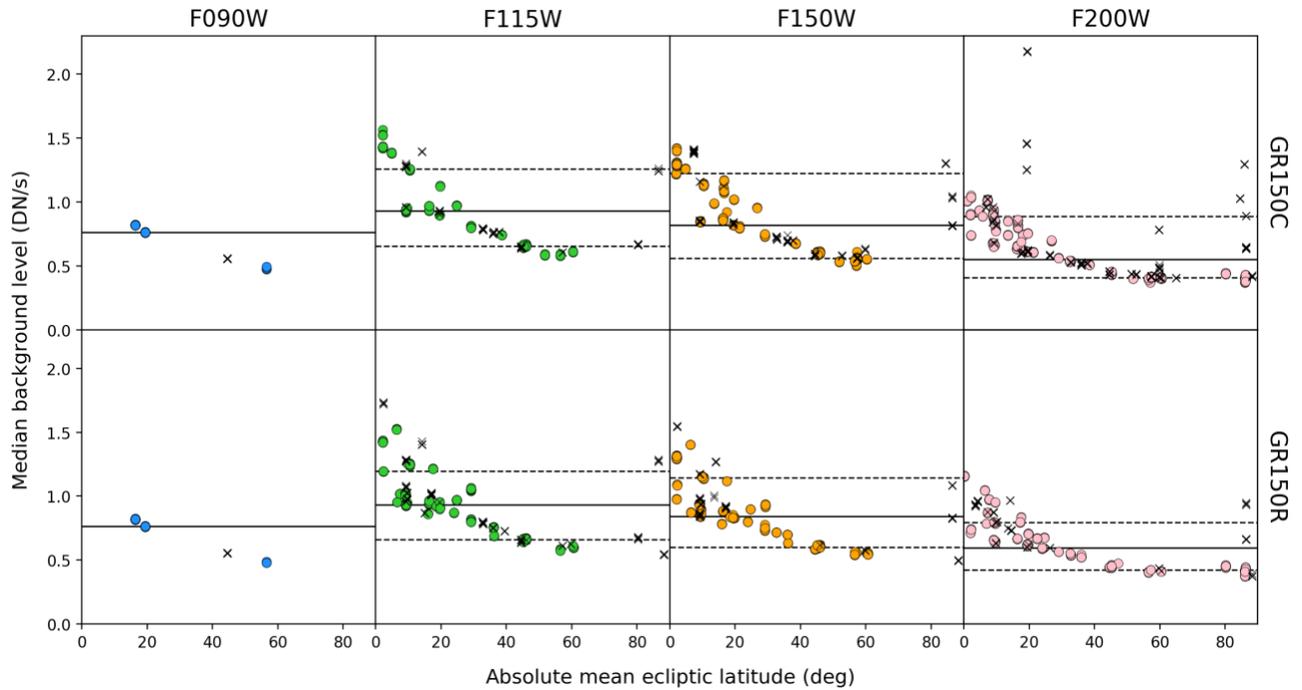

Figure 2. Median background level as a function of ecliptic latitude of the vetted (circles) and rejected (crosses) exposures inspected in this work. The horizontal solid lines indicate the median background level across all vetted exposures for each filter and grism pair, with the dashed lines indicating the 16th-84th percentiles (not shown for the F090W filter due to the small number of exposures). Overall, the background level decreases with higher ecliptic latitude and at redder wavelengths, which is expected if the signal is dominated by the zodiacal light. The median F090W background level will be reevaluated when more data become available.

artifacts and ensure that the background of the WFSS exposures is dominated by the sky signal.

- Next, we create a segmentation image of the spectral traces in each WFSS exposure using the `detect_source` function of the `photutils` python package (Bradley et al. 2025). For crowded extragalactic fields identified during the visual inspection, we use a more aggressive masking (i.e., a lower detection threshold) to ensure proper removal of the cluster light (e.g., intra-cluster light, or extended, bright ellipticals). The top panels of Figure 3 show an example WFSS exposure, its associated segmentation map, and the final masked image only leaving regions with sky background signal. The bottom panels of the Figure show our more aggressive masking applied to a crowded field, which successfully masks source spectra and the dispersed diffuse light in the field.

- The masked images are then median normalized to ensure that they are scaled to the same sky background level. Next, all WFSS exposures taken with the same filter and grism pair are stacked and the median sky intensity of each pixel through the stack is calculated, using a sigma-clipping outlier rejection (using 2.2 sigma and a maximum of 5 iterations). Additionally, we median normalize the resulting background image. This creates background models with median levels of 1 count rate (DN/s), useful to straightforwardly derive the sky level of science images during background scaling.

- Finally, we smooth-denoise the background images created for each grism and filter pair combination using the `Background2D` function of the `photutils` python package with a median background estimator using filter and box sizes of three-by-three pixels. Again, we median normalize the resulting background images to achieve median levels of 1 count rate (DN/s). Last, the flat-field is applied back to the background models to be compatible with the `jwst` pipeline[3] at the time of writing (version 1.18.0 or earlier), which first performs background subtraction and then flat-fielding.

Because of the sparse number of WFSS observations taken with the F090W filter to date, we adopt a slightly modified methodology to create the WFSS background models in that filter. After visual vetting, our images comprise 36 (35) F090W exposures crossed with the GR150C (GR150R) grism. The majority (32 science images) are of crowded galaxy cluster fields from PID 3362 (P.I.: Muzzin), while the remaining 4 (3) exposures are of a single sparse field from commissioning program 1089 (P.I.: Martel). Due to the low fraction of non-crowded WFSS vetted exposures for the F090W filter (~10%), we adopt a slightly more aggressive masking of the cluster fields compared to the other wide-band filters. This ensures the removal of the residual diffuse cluster light from the central area of the fields which would not be captured by the outlier rejection since the cluster fields make up the majority of the F090W exposures.

However, this creates higher pixel-to-pixel variations (i.e., higher noise) in the central area of the final F090W WFSS background images only covered by the small number of sparse exposures. Therefore, we also adopt a more aggressive smoothing of the final background images, using rectangular boxes of 9 pixels in the dispersion direction and 81 pixels in the cross-dispersion direction with a three-by-three median filtering to denoise the images. This strategy ensures that the smoothly varying background intensity is preserved while the noise for pixels with similar background level is efficiently reduced. This smoothing also removes the negative traces of the pick-off mirror occulting spots, which we extract and reinject using non-smoothed WFSS background images created from all F090W exposures.

---

[3] https://jwst-pipeline.readthedocs.io/

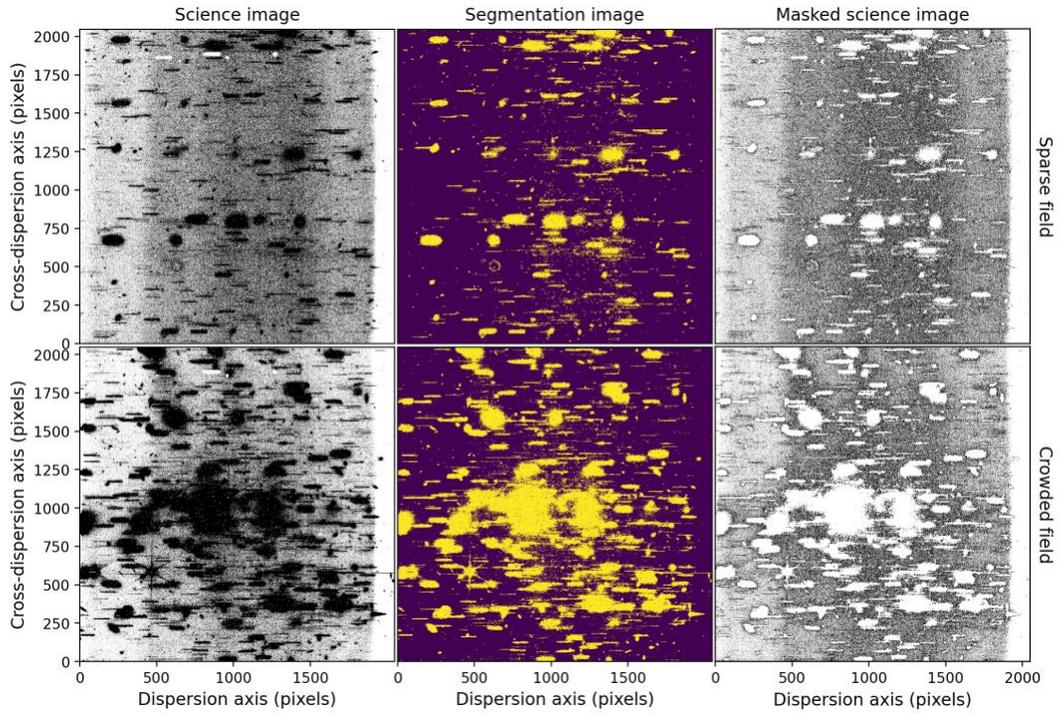

Figure 3. *Top*: An example WFSS science exposure, its associated segmentation map from the source detection step, and the resulting masked science image only leaving regions with background signal, from left to right, respectively. *Bottom*: same panels, but for a crowded field, where a slightly more aggressive source detection is applied.

## 3.2. Refined WFSS backgrounds

The background images created as described in Sec. 3.1 offer improved quality of background-subtracted science images compared to the initial set of background models created during commissioning with limited data (PIDs 1089 and 1448). The former are available on CRDS as of context 1307 to replace the commissioning images. However, as seen in Fig. 1, the NIRISS WFSS backgrounds have significant variations across the detectors, including sharp drops near the detector edges. In detector regions with low background signal, this results in some source pixels not being masked during our initial source detection and segmentation of the vetted exposures, while in high-background regions, this results in some background pixels having flux values above our detection thresholds and being masked during image segmentation. The former mostly affects areas near the detector edges where the background drops significantly, while the latter affects various areas with strong background signal such as the lightsaber and regions with significant overlap between the different sky spectral orders, including circular areas in the F115W GR150R and F200W GR150R backgrounds.

To mitigate this, we refine our final background images with an iterative approach similar to that of the HST WFC3/IR grism backgrounds (Kümmel et al. 2011). After the creation of our initial background images described in Sec. 3.1, we repeat our procedure starting at the source detection and segmentation step. However, we first background-subtract our vetted exposures using our background images created in Sec. 3.1. This creates WFSS exposures with relatively low background residuals across the detector, for which source detection and segmentation is improved, especially in the areas previously affected by the low or high background signals. We then apply all remaining steps as before (i.e., using our original vetted exposures), which creates new, improved background images due to the improved segmentation mapping. We repeat this procedure several times, each time using the newly created background images to background-subtract the vetted exposures and create improved segmentation maps in the next iteration. We stop after five iterations when there is no more improvement of the final models. This iterative procedure is applied to the F115W, F150W, and F200W filters crossed with both grisms. The top panel in Figure 4 shows the initial and final segmentation maps obtained for an example exposure in the F150W GR150C affected by strong background, as well as their difference image. The bottom panel of the Figure shows the initial and final background images in that filter and grism pair and the fractional difference between the two. The difference images particularly reveal the regions of high backgrounds (due to the overlap of the sky spectral orders) recovered in our refined, final models. We do not apply this procedure to the F090W filter, but will do so as more F090W exposures become available. These refined backgrounds are available on CRDS as of context 1365.

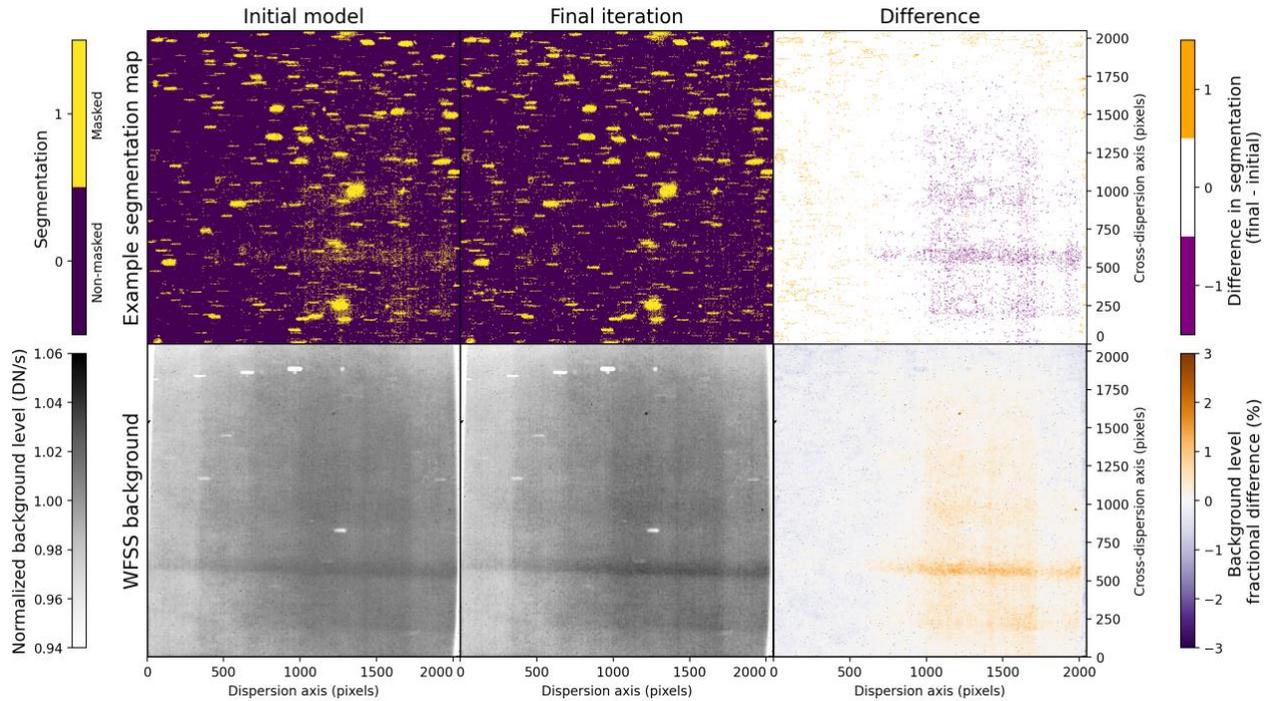

Figure 4. Comparison between initial and final iterations of the refined modelling. *Top*: Segmentation map of an example F150W/GR150C exposure, where the left panel shows the initial map, the middle panel shows the final map derived in the last iteration of our refined modelling, and the right panel shows the difference between the two. Pixels with high-background signal that are initially masked and successfully recovered in the last iteration are visible in purple in the right panel. The difference image also reveals trace residuals (orange pixels) from low-background regions that are missing from the initial segmentation but successfully masked in the last iteration. *Bottom*: Initial F150W/GR150C background, final background after the last iteration, and fractional difference between the two, from left to right, respectively. As seen in the difference image, the final model, compared to the initial model, has a higher background level in the regions that are no longer systematically masked and a slightly lower background level in the regions that have improved trace masking.

### 4. Quality of the WFSS backgrounds

Figure 5 shows our final WFSS background models after flat-fielding for all filter and grism combinations. Each panel shows the 2D background image and median 1D profiles in each axis. The most apparent features in the 2D backgrounds are the lightsabers at around 500 pixels in the Y-axis, the negative traces due to the pick-off mirror occulting spots around 1800 pixels in the Y-axis, and the overlap of the different spectral orders of the sky background with respect to the dispersion direction (X-axis for the GR150C and Y-axis for the GR150R grisms). The latter is most prominent in the reddest filters, especially the F200W/GR150C and F200W/GR150R cases. As shown in Fig. 2, they are consistent with having the lowest zodiacal contribution among all filters, which is expected from the monotonically decreasing spectrum of the zodiacal light between ~1 and 3 μm. Note, however, that the sky background level is also affected by stray-light and other components, and does not exactly follow model predictions especially at the shortest wavelengths (Rigby et al. 2023b)[4]. The stronger variation of the background for the redder filters is the result of the longer traces and gaps between orders at longer wavelengths. As expected, the 1D profiles of the WFSS backgrounds are relatively flat with respect to the cross-dispersion direction, except around 500 pixels for the GR150C due to the lightsaber contribution. This contribution is seen with respect to the dispersion-direction for the GR150R backgrounds, and appears blended with the overlap of the spectral orders in the F200W/GR150R background. On the other hand, the lightsaber does not appear as a feature of the F090W backgrounds. We interpret this as the result of the very small number of non-crowded, vetted exposures used to create our F090W backgrounds, and we will revisit this interpretation as more F090W exposures become available.

The new WFSS backgrounds offer substantial improvements compared to the commissioning models derived using limited data. In the following, we refer to the commissioning models as the "previous backgrounds". Figure 6 shows two example science exposures from program 1571 (P.I.: Malkan) taken with the F200W filter crossed with the GR150C and GR150R grisms, respectively. The 2D panels show the residuals after background subtraction and flat-fielding when using the previous (top) and new (bottom) backgrounds. As seen in the Figure, significant portions of the images suffer from over- or under-subtraction of the background by up to 2-5% of the background level, when using the previous models. In comparison, the 2D residuals using the new backgrounds are well distributed around zero with reduced global scatter (< 2%). This is further seen in the bottom panel of the Figure, which shows the median residual profiles with respect to the dispersion direction for each case. While the residual background level varies significantly (up to 5%) across the dispersion axis when using the previous backgrounds, it remains very well under the 1% level (dashed lines) when using the new backgrounds. Hviding et al. (2024) independently derive improved WFSS backgrounds compared to the commissioning models, using a strategy roughly similar to ours. We provide a comparison to their models and our new WFSS backgrounds in Appendix A. In short, their models (which do not include the F090W filter) are similar to ours down to the 1% level, with the largest discrepancies in the F115W/GR150C, where their model seems affected by some features resembling stars and diffraction spikes.

---

[4] Also see Martel et al. (in prep) for a detailed analysis of the NIRISS imaging background.

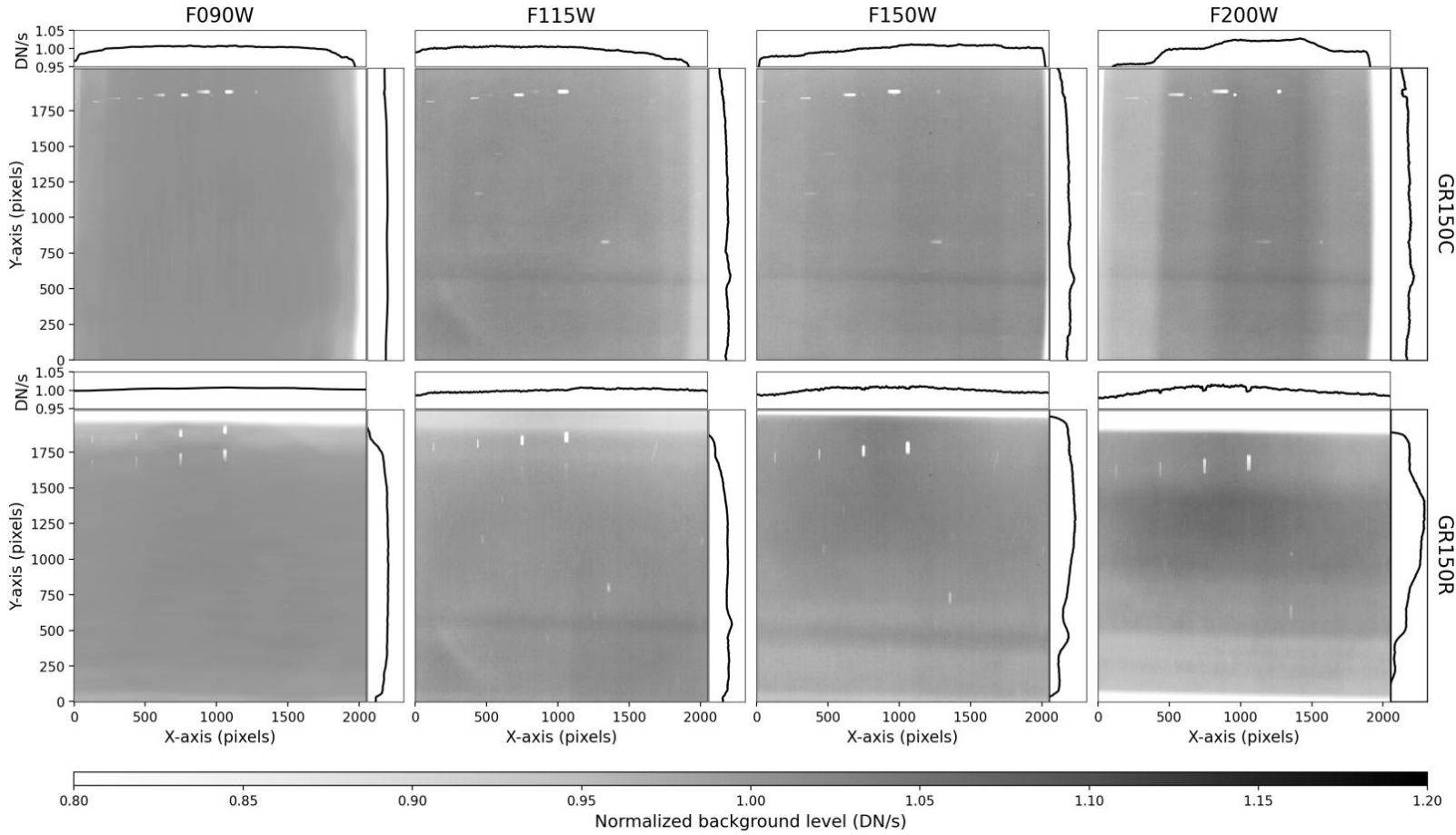

Figure 5. Final WFSS background images after flat-fielding for all filter and grism combinations (top: GR150C, bottom: GR150R). Each panel shows the median normalized background (in count rates) after flat-fielding applied. The median 1D profile along each axis is also shown on the sides of each panel. The varying intensity and different shapes of the background signal across the detector for the different filter and grism combinations is clearly seen in the 2D images and 1D profiles.

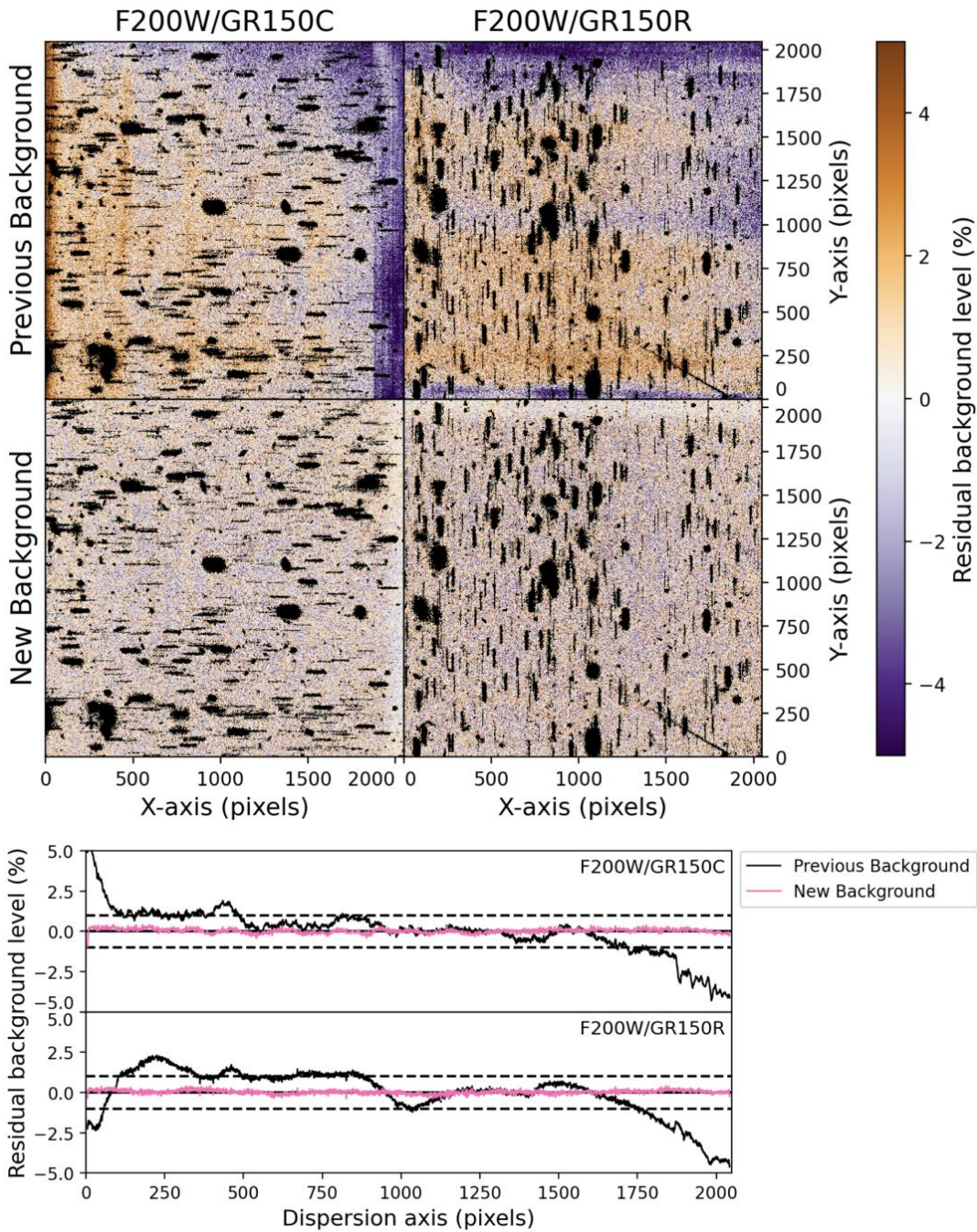

Figure 6. *Top*: 2D background residuals on two F200W example exposures (left: GR150C, right: GR150R) after background subtraction, using the commissioning background models (top, "previous" background), and our new backgrounds (bottom). The background-subtracted exposures show strong artifacts in the top panels, which are completely corrected when using our new models. *Bottom*: median 1D profiles along the dispersion axis of the images in the top panel, derived from regions without traces. The pink (black) curves show the background residuals when using the new (previous) backgrounds. The horizontal dashed lines indicate the $\pm 1\%$ level. The residual background level using our new backgrounds is well under 1% of the sky brightness.

To quantitatively assess the improvements offered by our new models, we derive the background residuals after background subtraction and flat-fielding for a relevant number of exposures, using both the previous and new WFSS backgrounds. To avoid circular comparisons we select relatively small, but relevant, sub-samples of images, consisting of 15 randomly selected exposures for the F090W/GR150C and F090W/GR150R, and 31 for the other filters for which the parent samples of vetted exposures are much larger (cf. Table 1). We then perform the background subtraction of all randomly-selected exposures using the same scaling method[5] as of `jwst` pipeline version 1.17.0 with the exception of using our own trace masking (see Sec. 3.1 and 3.2), and apply the flat-fielding. The scaling method computes the inverse-variance weighted mean of the ratio between the data and the background model, with an iterative outlier rejection parametrized by the maximum number of iterations, the rejected outlier fraction, and an RMS stopping criterion. We use 20 maximum iterations, a 2% outlier rejection (keeping 98% of the data), and a stopping RMS threshold of 1% between iterations. For all filter and grism pairs, we then derive the median 1D profiles of the residual background levels of all exposures from regions devoid of spectral traces. Figure 7 shows the 1D profiles derived along the dispersion direction and cross-dispersion direction for all filter and grism pairs. The black and pink curves indicate the residual background levels when using the previous and new backgrounds, respectively. The new F090W backgrounds show the largest improvements across all filters and grism pairs compared to the previous models. This is due to the very small number of exposures (< 5) available in that filter for each grism during commissioning, resulting in residuals showing variations of up to 5% of the background level across significant portions of the detector when using the previous backgrounds. In comparison, our new backgrounds, while still derived using a small number of sparse fields together with highly-crowded fields, only show small variations in the residual profiles, remaining under the 1% level. The previous backgrounds also suffer from non-negligible trends for the other filter and grism pairs with respect to both axes, except the F115W filter which shows relatively flat residuals using the previous backgrounds (albeit with some small variations and trends reaching the 1% level). As seen in the Figure, our new WFSS backgrounds do not suffer from these trends and show flat residuals in both the dispersion and cross-dispersion directions, with median residual profiles centered on zero and variations well under the 1% level (dashed lines).

We further quantify the overall improvement between the previous and new WFSS backgrounds by deriving the RMS of the median residual profiles along the dispersion direction (in percentage of the background level) from the same set of random exposures as before. The top (bottom) panel in Figure 8 shows the RMS values for all filters in the GR150C (GR150R) grism, where square symbols represent the RMS derived from the previous backgrounds, and the circles the RMS derived from the new backgrounds. As seen in the Figure, the RMS for the F090W filter crossed with either of the grisms decreases by a factor of ~7, from about 3% to 0.4% of the background level, which is the largest improvement among all WFSS backgrounds. For the F115W, F150W, and F200W filters crossed with the GR150C grism, the RMS improves by a factor of 1.1, 2.0, and 2.1, respectively, while it improves by a factor of 1.8, 3.4, and 3.4, for the F115W, F150W, and F200W filters crossed with the GR150R grism. Overall, the RMS of the new WFSS

---

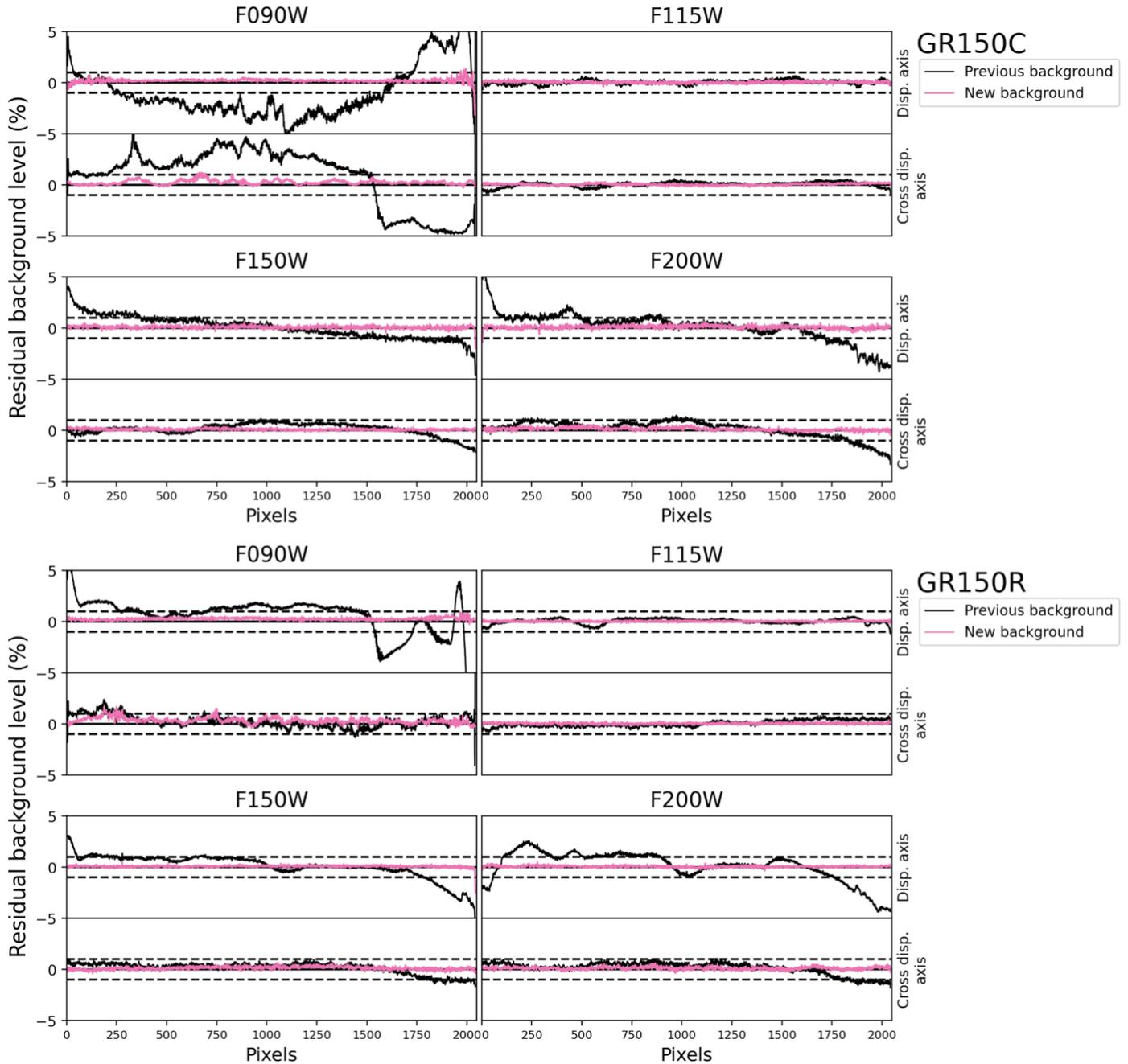

Figure 7. Same as the bottom panel of Fig. 6, but derived using a statistical number of background-subtracted exposures, and shown for all filter and grism combinations (top grid: GR150C, bottom grid: GR150R). For each filter and grism panel, the top shows the 1D median profiles along the dispersion direction, and the bottom shows the profiles along the cross-dispersion direction. As seen, the strong background residuals affecting background-subtracted exposures when using the previous models (black curves) are completed corrected when using our new models (pink curves), and any remaining residuals are under the 1% level (dashed lines).

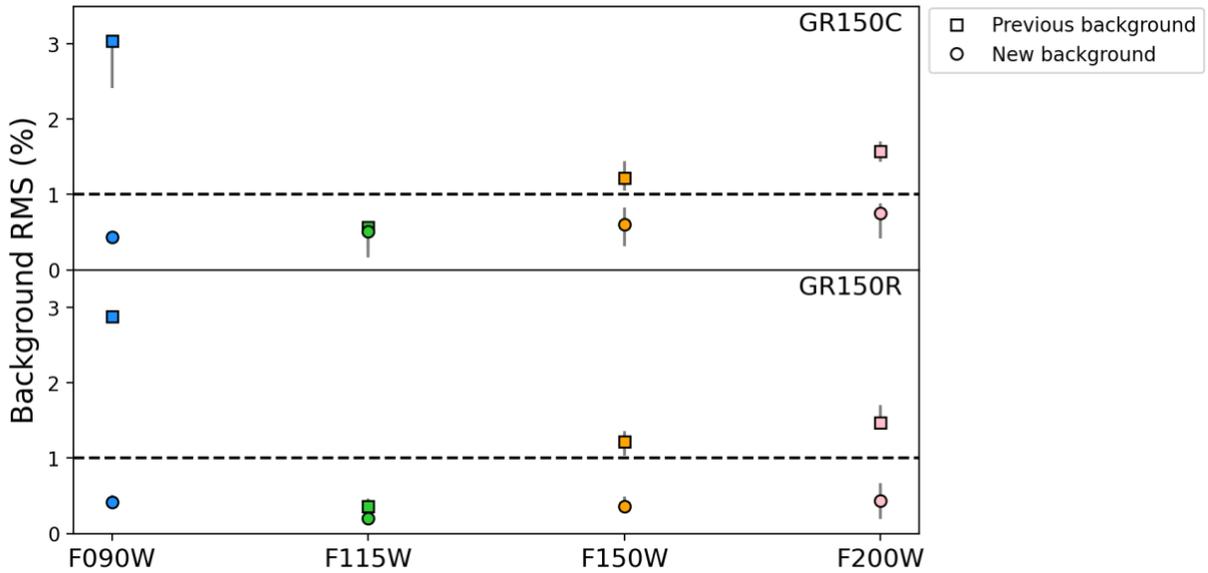

Figure 8. Root-mean square (RMS) of the median residual profiles (along the dispersion direction) derived from a statistical number of background-subtracted exposures, for all filter and grism combinations (top: GR150C, bottom: GR150R). The RMS is shown in percentage of the sky brightness, and circles (squares) correspond to using the new (previous) WFSS backgrounds for the background-subtraction. The dashed lines indicate the 1% level. As seen, the RMS is below the 1% level for all filter and grism combinations when using our new backgrounds.

backgrounds reaches below the 1% level for all filter and grism combinations, and is below the 0.5% level for all barring the F150W/GR150C and F200W/GR150C which have RMS of 0.60% and 0.75%, respectively.

## 5. Conclusions

We present new JWST/NIRISS background models for the Wide-Field Slitless Spectroscopic mode of observation for the F090W, F115W, F150W, and F200W filters crossed with the two orthogonal grisms (GR150C, and GR150R). These models are derived using carefully vetted science and calibration exposures taken between 2022 May 05 and 2024 June 11 and improve the residuals of background-subtracted science images compared to previous generation of models, for all filter and grism combinations. Overall, the residuals of background-subtracted images are well-behaved and centered on zero with scatter well below the 1% level. The new models correct the strong variations and trends affecting the previous WFSS backgrounds derived during commissioning which suffer from artifacts at the 1-5% level depending on filter and grism combinations. The RMS of background residuals improve by a factor of ~7 for the F090W backgrounds and factors of 1.1-3.4 for the other filters, reaching RMS values below the 1% level. These new, empirical WFSS backgrounds are available on CRDS as of context 1365, and we recommend using them to perform the background-subtraction of NIRISS WFSS science images taken with the F090W, F115W, F150W, or F200W filters crossed with the GR150C or GR150R grisms.

While the new models offer significant improvements over previous WFSS backgrounds, we note that the F090W backgrounds are derived using a limited number of images. Calibration program 9286 (P.I.: Noirot) will obtain additional observations with the GR150C and GR150R grisms crossed with the F090W filter, which we will use to upgrade our models as the data become available. Additionally, bright stars falling in the susceptibility region might increase the lightsaber signal to ~10% of the background level, compared to a few percent when the lightsaber is dominated by the zodiacal light (Rigby et al. 2023a). While approved programs are checked during scheduling to avoid this issue (when desired), we note that an independent modelling of the lightsaber could allow better removal of residual lightsaber light not captured by our models.

### Acknowledgments
GN acknowledges support by the Canadian Space Agency under a contract with NRC Herzberg Astronomy and Astrophysics. This work is based on observations made with the NASA/ESA/CSA James Webb Space Telescope. The data were obtained from the Mikulski Archive for Space Telescopes at the Space Telescope Science Institute, which is operated by the Association of Universities for Research in Astronomy, Inc., under NASA contract NAS 5-03127 for JWST. GN would like to thank the wonderful NIRISS instrument team he is part of, and in particular Paul Goudfrooij, Stephanie LaMassa, Rachel Plesha, Jo Taylor, Kevin Volk, and Chris Willott, for useful inputs and discussions regarding this work.
Facilities: JWST (NIRISS)
Software: astropy (Astropy Collaboration et al. 2022), matplotlib (Hunter 2007),

numpy (Harris et al. 2020), photutils (Bradley 2025).

## A. Comparison to the literature

Hviding et al. (2024; hereafter HMC24) derived WFSS backgrounds for the F115W, F150W and F200W filters crossed with the two orthogonal grisms, as well as NIRISS imaging backgrounds in the same wide-band filters. The methodology they use to create the WFSS backgrounds is roughly similar to ours. In short, they first query MAST to retrieve all public WFSS data as of 2024 September 24 in the appropriate filters, and exclude exposures affected by bright nearby stars based on a Gaia DR3 search (Gaia Collaboration et al. 2023). However, they do not perform a visual vetting of the exposures and also include data with high background level such as the Large Magellanic Cloud and M33 which are removed in our approach (see Fig. 2, and Fig. 2 in HMC24). Then, they use the `Source Extractor` (Bertin & Arnouts 1996) python wrapper sep (Barbary 2016) to model the 2D background of each exposure, and use `photutils` to perform source detection and masking with a slightly higher detection threshold than in our work. This threshold is applied to all exposures, whereas we also apply slightly more aggressive thresholds to crowded fields compared to non-crowded ones. Subsequently, they median-combine the masked exposures similarly to our method, and fill empty pixels with `maskfill` (van Dokkum & Pasha 2024). The resulting models are then smoothed using a non-local filtering method (Darbon et al. 2008, van der Walt et al. 2014) and the process is iterated using a similar approach as ours. They lower the detection threshold during the additional iteration and perform the iterative process a single time, which they find is sufficient to converge to a final background with their method.

Figure A1 shows the comparison between the HMC24 backgrounds[6] and the ones we derive, excluding the F090W WFSS backgrounds which are not included in HMC24. The top grid shows the GR150C backgrounds and the bottom grid the GR150R backgrounds. In each grid, the top panels are the HMC24 models, the middle panels are the new models derived in this work, and the bottom panels show the fractional difference between the two. Overall, the backgrounds appear very similar down to the 1% level, with most apparent background features (i.e., the overlap between the different spectral orders, and the lightsaber) showing no or little differences. Across all filter and grism combinations, the main differences are seen around the negative occulting-spot traces and towards the detector edges with respect to the dispersion direction. The former might be caused by the slightly different smoothing algorithm used between the two methodologies, and the latter might be the result of the sharp decrease in background signal towards the detector edges, creating large fractional differences even for small variations between the models. We also note the presence of what appears to be faint flat-field residuals across all of HMC24 backgrounds (roughly between 1000 and 2000 pixels along the X-axis and between 1000 and 1500 pixels along the Y-axis), corresponding to the contours of the low-epoxy region in NIRISS flat-fields[7]. These residuals may be due to improper flat-fielding of some exposures, an effect of the various rescaling of the exposures during the creation of the backgrounds, or an evolution of the low-epoxy edges over time which could appear in HMC24 models due to their different smoothing strategy. Precisely determining the cause of these residuals in HMC24 models is beyond the scope of this work; we simply caution the reader of these differences between the models. Last, we note additional differences mostly located in the bottom and upper right corners of the F115W/GR150C background between the HMC24 and our new models. These appear as features resembling stars and diffraction spikes in the HMC24 model, which are visible in the difference image. We speculate that these are due to improperly masked stars and diffraction spikes in some HMC24 F115W/GR150C exposures. We attribute other differences seen below the 1% level, such as large-scale variations between the models, as the result of the slight differences between the two methodologies (e.g., exposures used, source detection and masking, smoothing, scaling method, etc.), and we remind the reader that the HMC24 WFSS backgrounds and ours are overall consistent down to the 1% level.

---

[6] We download the latest available backgrounds at the time of writing, which are version 5 of the backgrounds published on 2024 September 24, and available at: https://doi.org/10.5281/zenodo.13838016

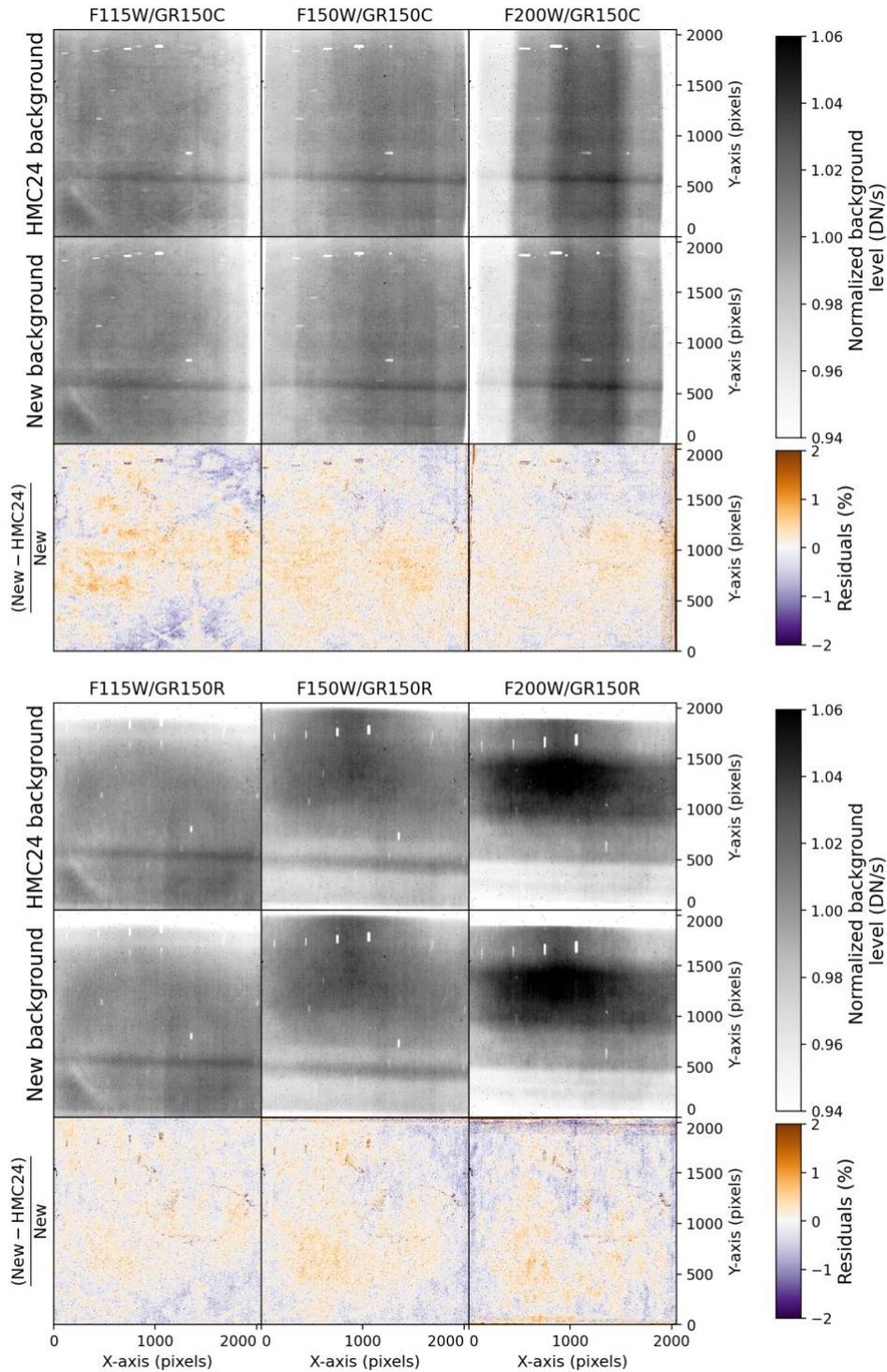

Figure A1. Comparison between the WFSS backgrounds derived in HMC24 and the ones derived in our work. The top (bottom) grid shows the GR150C (GR150R) backgrounds. Each grid shows the HMC24 backgrounds, our new backgrounds, and the fractional difference between the two, from top to bottom, respectively. Overall, the backgrounds are consistent down to the 1% level, with the strongest differences seen in the F115W/GR150C (potential star and diffraction spike residuals in the HMC24 background), and around the negative occulting-spot traces and the detector edges for all models.